\documentclass[reprint,amsmath,amssymb,aps,nofootinbib]{revtex4-1}

\usepackage{graphicx}
\usepackage{dcolumn}
\usepackage{bm}
\usepackage[normalem]{ulem}

\usepackage{color}
\usepackage{xcolor}

\usepackage{mathrsfs}
\usepackage{amsfonts}
\usepackage{varioref}
\usepackage{csquotes}
\usepackage{hyphenat}
\usepackage{float}
\usepackage{multirow}

\RequirePackage[colorlinks,citecolor=blue,urlcolor=magenta,linkcolor=blue]{hyperref}

\def\KN{Kerr-Newmann }

\def\KN{Kerr-NUT }

\labelformat{chapter}{Chapter (#1)} 
\labelformat{section}{Section (#1)} 
\labelformat{subsection}{Section (#1)} 
\labelformat{subsubsection}{Section (#1)}
\labelformat{subsubsubsection}{Section (#1)}
\labelformat{equation}{Eq.~(#1)} 
\labelformat{figure}{Fig.~(#1)} 
\labelformat{subfigure}{Fig.~(\thefigure#1)} 
\labelformat{table}{Tab.~(#1)} 
\labelformat{appendix}{Appendix #1}

\def\ES {equatorial reflection symmetry}

\def \KN {Kerr-NUT}

\begin{document}

\title{Possible connection between the reflection symmetry and existence of equatorial circular orbit}

\author{Sayak Datta} 
\email{skdatta@iucaa.in}

\author{Sajal Mukherjee}
\email{sajal@iucaa.in}
\affiliation{Inter-University Centre for Astronomy and Astrophysics, Post Bag 4, Pune-411007, India}

\begin{abstract}
We study a viable connection between the circular-equatorial orbits and reflection symmetry across the equatorial plane of a vacuum stationary axis-symmetric spacetime in general relativity. The behavior of the circular equatorial orbits in the direction perpendicular to the equatorial plane is studied, and different outcomes in the presence and in the absence of the reflection symmetry are discussed. We conclude that in the absence of the equatorial reflection symmetry neither stable nor unstable circular orbit can exist on the equatorial plane. Moreover, to address the observational aspects, we provide two possible examples relating gravitational wave astronomy and the thin accretion disk which can put constraints on the symmetry breaking parameters.
\end{abstract}

\maketitle

\section{Introduction}

The Kerr metric uniquely describes a stationary, axis-symmetric, and asymptotically flat black hole (BH) solution of vacuum Einstein's field equations in four dimensions (assuming the regularity on and outside of the horizon) \cite{Kerr:1963ud, Israel:1967wq, Wald:1971iw, Carter:1971zc, Robinson:1975bv}. Besides the stationary and axis-symmetry properties, Kerr spacetime is also endowed with an additional feature of reflection symmetry across the equatorial plane. Even if the former characteristics are likely to be associated with astrophysical objects with rotation, both asymptotic flatness and equatorial reflection symmetry can be relaxed in order to probe a larger domain of compact objects other than BH. Various possible distinctions between BHs and other exotic compact objects based on tidal deformability \cite{Cardoso:2017cfl, Sennett:2017etc, Maselli:2017cmm, Brustein:2020tpg}, tidal heating \cite{Datta:2019euh, Maselli:2017cmm, Datta:2019epe, Datta:2020gem, Datta:2020rvo}, multipole moments \cite{Krishnendu:2017shb,Datta:2019euh}, {\it echoes} in postmerger \cite{Cardoso:2016rao, Cardoso:2016oxy, Tsang:2019zra,Abedi:2016hgu, Westerweck:2017hus, Cardoso:2019rvt} and electromagnetic observations \cite{Titarchuk:2005rr,Bambi:2013sha,Jiang:2014loa,Bambi:2015kza,Bambi:2015kza, Cardoso:2019rvt} have been proposed in the literature. Similarly, distinguishing them on the basis of \ES~can be useful to detect them or rule them out as viable astrophysical bodies. Not only may these studies provide a fresh outlook to model astrophysical objects, but they also may assign an observational impact to it. In the present article, we aim to elaborate on the equatorial reflection symmetry in a generic spacetime and outline its possible theoretical and observational implications in depth.

To study any particular effect appearing from spacetime geometry, the ideal approach is, to begin with, the orbital dynamics. Based on how orbits behave in a given spacetime, more involved astrophysical searches are constructed. In Kerr, the orbital properties are well studied and extensively explored in literature \cite{Wilkins:1972rs,o2014geometry,chandrasekhar1998mathematical}. Similar exploration is carried out for Kerr-NUT spacetime \cite{Jefremov:2016dpi,Mukherjee:2018dmm,Chakraborty:2019rna}, which violates the \ES~and describes an asymptotically nonflat geometry \cite{Newman:1963yy,LyndenBell:1996xj}. While in Kerr we know stable/unstable equatorial circular orbits exist, the same is not true in the presence of NUT charge. In particular, neither stable nor unstable equatorial circular orbits can exist for massive or massless particles in Kerr-NUT spacetime \cite{Jefremov:2016dpi, Mukherjee:2018dmm}. In fact, this stark contrast between Kerr and Kerr-NUT is the primary source of our motivation to study further and investigate whether \ES~and the existence of the equatorial circular orbits can be generically connected. We introduce a perturbative approach for confronting equatorial circular geodesics in geometries where \ES~is absent. By assuming that the orbits reside on the equatorial plane initially, we study the growth of perturbation in time and aim to realize which parameters engineer any possible deviation from the equatorial plane. Assuming the perpendicular to $\theta=\pi/2$ is along the z axis, we consider the z perturbation in our work. Because of the absence of the \ES, it is likely that the potential will not be an even function of z. As a result, there will be an intrinsic force in the z direction that will lift the orbits from the equatorial plane. To study this in the context of a general stationary axis-symmetric metric we will consider Ernst's potential and write metric components in terms of it.

The existence of the planner circular orbits is crucial for discussing physical effects relevant for various astrophysical models, such as binary, spectra of accreting black holes, etc. The orbits in extreme mass ratio inspirals, which will be observed with LISA \cite{Audley:2017drz}, are most likely to be generic \cite{Amaro-Seoane:2014ela, Barack:2003fp, Merritt:2011ve}. Binaries with stellar masses, as already detected by gravitational wave detectors LIGO and VIRGO, can have precession \cite{Apostolatos:1994mx}. This means that the understandings found in the current paper will not only have theoretical grounds but also it will have an observational impact, which will be discussed later. 

The rest of the manuscript is organized as follows. In \ref{Ernst}, we start with the Ernst potential, and in \ref{Geodesic}, we will briefly discuss the geodesic equations in terms of the metric components. The primary findings of the paper are given in \ref{decomposition} and \ref{results}, respectively. In \ref{observation}, we will study the observational impacts of current findings, and finally, we will conclude in \ref{conclusion}.


\section{Metric components and the Ernst potential}
\label{Ernst}

We start with a stationary, axis-symmetric, and vacuum spacetime written in cylindrical coordinates $(t,\rho,z,\phi)$ within general relativity as \cite{Ryan:1995wh, wald2010general}
\begin{equation}
\label{metric}
    ds^2 = -F(dt -\omega d\phi)^2 + \frac{1}{F}\big[e^{2\gamma}(d\rho^2 + dz^2) + \rho^2d\phi^2\big],
\end{equation}
where $F$, $\omega$, and $\gamma$ are functions of $\rho$ and $z$. Substituting the metric in the Einstein equation, it is possible to find governing equations for these entities. In passing, we should note that the above metric does not guarantee to be asymptotically flat and remains general otherwise. 

The above metric components can be written in a more compact form by using the complex Ernst potential, which is a combination of both a \textit{norm} ($\lambda$) and \textit{twist} ($\omega$) timelike Killing vector. In particular, $\lambda=-g_{tt}=F$, and $\omega_{\mu}=\sqrt{-g}\epsilon_{\mu \nu \gamma \delta} \xi^{\nu}\nabla^{\gamma}\xi^{\delta}$, where $\xi^{\mu}$ is the timelike killing vector \cite{wald2010general}. Given that the spacetime is stationary and axis symmetric, both norm and twist are expected to be nonzero. Finally, the Ernst potential takes the form
\begin{equation}
    \mathcal{E} = F + i\psi = \frac{(\rho^2 + z^2)^{1/2} - \Tilde{\xi}}{(\rho^2 + z^2)^{1/2} + \Tilde{\xi}},
\end{equation}
where $\Tilde{\xi}$ can be written as \cite{fodor1989multipole}
\begin{equation}
    \Tilde{\xi} = \sum_{j,k = 0}^{\infty} a_{jk} \frac{\rho^j z^k}{(\rho^2 + z^2)^{j+k}}.
    \label{eq:ajk}
\end{equation}
The reasons to choose Ernst's potential formalism as a tool to express metric components are twofold. First, the metric components can be written directly in terms of $\mathcal{E}$. As a result, from the behavior of $\mathcal{E}$ under the absence of the symmetries, the nature of the orbits can easily be extracted. Second, reflection symmetry manifests itself through the values of $a_{jk}$, making it easier to impose the presence or the absence of \ES. The $a_{jk}$ is nonzero only for non-negative, even $j$ and non-negative $k$. If there is reflection symmetry across the equatorial plane, then $a_{jk}$ is real for even $k$ and imaginary for odd $k$ \cite{Ryan:1995wh,fodor1989multipole,Ernst:2006yg,Ernst:2007xq}. However, as the present study remains general as far as the \ES~is concerned, we restrain ourselves to make such assumptions. We assume that $a_{jk}$ has both real and imaginary components for both even and odd values of $k$.

In terms of $F$ and $\psi$, the metric components $g_{tt}$, $g_{t\phi}$ and $g_{\phi \phi}$, which will be of particular use, can be written as \cite{Ryan:1995wh}
\begin{eqnarray}
\label{eq:g_tPhi}
   g_{tt} &=& -F, \quad g_{\phi \phi}=(g_{t\phi}^2-\rho^2)/g_{tt}, \nonumber \\
     g_{t\phi} &=& -F \int \limits_{z=const} \frac{\rho}{F^2} \frac{\partial \psi}{\partial z} d\rho +F \int \limits_{\rho=const} \frac{\rho}{F^2} \frac{\partial \psi}{\partial \rho} dz,
  \end{eqnarray}
where the detailed calculations to arrive at the following expressions are shown in \ref{Ernst_Pot}.


\section{ Off-equatorial perturbation of the equatorial geodesics}
\label{Geodesic}
To study the existence of equatorial circular orbits in a generic spacetime with metric given in \ref{metric}, we start with the geodesic equations
\begin{eqnarray}
\Ddot{\rho} =\frac{1}{2}\frac{\partial \mathcal{V}_{\rho}}{\partial \rho}, \quad \Ddot{z} =\frac{1}{2}\frac{\partial \mathcal{V}_{\rm z}}{\partial z},
\label{eq:geodesic}
\end{eqnarray}
where the dot defines a derivative with respect to the affine parameter which we may call $\tau$, and $\mathcal{V}_{\rho}$ and $\mathcal{V}_{z}$ are radial and angular potentials, respectively. The above equations will determine the locations of an orbit while the conserved energy and momentum are dictated by $t$ and $\phi$ components. Given that we are interested in orbits confined on a plane and circular in nature, the above two equations would give $\Ddot{\rho}=\Ddot{z}=0$ in principle. However, as the \ES~is not respected, the potential $\mathcal{V}_{z}$ is likely to contain terms with the odd power of $z$ such that $\mathcal{V}_{z}(z) \neq \mathcal{V}_{z}(-z)$. One should be mindful that the condition of circularity is not expected to be affected by the \ES, and we may safely impose $\dot{\rho}=\Ddot{\rho}=0$. From the timelike constraint, $\mathcal{U}^{\alpha}\mathcal{U}_{\alpha}=-1$, we arrive at the expression
\begin{equation}
    \mathcal{U}^{t} \mathcal{U}_{t}+\mathcal{U}^{\rho}\mathcal{U}_{\rho}+\mathcal{U}^{z} \mathcal{U}_z+\mathcal{U}^{\phi}\mathcal{U}_{\phi}=-1,
\end{equation}
and finally \cite{Mino:2003yg}
\begin{eqnarray}
\mathcal{V}_{z}=g_{zz}(\mathcal{U}^z)^2=-1-g^{tt}E^2-g^{\phi \phi}L_{\rm z}^2-2 g^{t\phi}EL_{\rm z},
\label{eq:Vz}
\end{eqnarray}
where $E$ and $L_{\rm z}$ are given as conserved energy and momentum respectively, appearing from the spacetime symmetries. We will expand $\mathcal{V}_{z}$ about the equatorial plane, i.e., $z=0$. Neglecting terms $\sim \mathcal{O}(\delta z)^3$ and beyond, i.e., $\mathcal{V}_{z}(\delta z)=\mathcal{V}_{0}+(\delta z) \mathcal{V}_{1}+(\delta z)^2 \mathcal{V}_{2}$, we may rewrite z equation in \ref{eq:geodesic} as follows:
\begin{eqnarray}
2\Ddot{\delta z}=\mathcal{V}_{1}+2 \mathcal{V}_{2} \delta z.
\label{eq:pot_eq}
\end{eqnarray}
By solving the above equation, we arrive at
\begin{equation}
\delta z(\tau) = \dfrac{\mathcal{V}_1}{2\omega^2}+A \exp[-i \omega \tau]+B \exp[i \omega \tau],    
\end{equation}
where we set $ \mathcal{V}_{2}=-\omega^2$, and $\omega$ can be both real and imaginary. If we assume that the initial conditions are, $\delta z(0)=\dot{\delta z}(0)=0$, the above equation may be rewritten as
\begin{equation}
\label{eq:bc_solution}
\delta z(\tau) = -\dfrac{\mathcal{V}_1}{2\omega^2} +\dfrac{\mathcal{V}_1}{4 \omega^2}  \left\{\exp(-i \omega \tau)+\exp(i \omega \tau)\right \}. 
\end{equation}
Depending on the nature of $\omega$, the above equation either represent a hyperbola ($\rm{Im}[\omega] \neq 0$), and a oscillatory ($\rm{Im}[\omega] = 0$) motion. For the Kerr-NUT spacetime, as we have shown in \ref{Appen:Kerr_NUT}, $\rm{Im}[\omega] = 0$, and the solution is always oscillatory. For future purposes, we may note that in case of a oscillatory solution we have
\begin{equation}
\delta z(\tau)=-\dfrac{\mathcal{V}_{1}}{\omega^2} \sin^2(\omega \tau/2),
\label{eq:sintheta}
\end{equation}
which hints at an interesting property of this motion written as follows. For a particular radius, this perturbation is either positive or negative depending on the sign of $\mathcal{V}_1$, but never switches sign. It indicates that it would never cross the equatorial plane but approach it in each cycle. In short, the hobbling from the equatorial plane would be one sided. 

It should be mentioned that the above equation built in with a condition where there is no external perturbation otherwise $\delta z(0) \neq 0$. Any additional external perturbation may result in some changes in the final expression which we have not studied in this article. Besides, one should also note how $\mathcal{V}_{1}$ and $\mathcal{V}_{2}$ are affecting the perturbation. While $\mathcal{V}_{1}$ is directly proportional to its value, $\mathcal{V}_{2}$ is responsible for engineering its nature. By setting $\mathcal{V}_{2}=0$, we obtain a diverging nature of the perturbation, as $\delta z(\tau) \sim \tau^2$, and may not appropriately model $\delta z(\tau)$. Keeping these points in mind, we will keep $\mathcal{V}_1$ and $\mathcal{V}_2$ in the expression and ignore higher-order corrections. Eventually, we will show that nonzero $\mathcal{V}_1$ is connected to the absence of \ES, which, as a result, does not allow a circular orbit to exist in the equatorial plane, not even perturbatively.

\section{Decomposition of the Ernst potential}
\label{decomposition}
Referring to \ref{eq:ajk}, we may state that \ES~in the potential comes through $a_{jk}$. For a clear exposition of our results, we separate out the real and imaginary parts of $a_{jk}$, i.e., 
\begin{equation}
    a_{jk} = \hat{a}_{jk} + i \Breve{a}_{jk},
\end{equation}
where $\hat{a}_{jk}$ and $\Breve{a}_{jk}$ are the real and imaginary parts of the $a_{jk}$, respectively. If the \ES~exists, then $a_{jk}$ is real for even $k$ and imaginary for odd $k$ \cite{Ryan:1995wh,fodor1989multipole,Ernst:2006yg,Ernst:2007xq},
\begin{eqnarray}
\Breve{a}_{j(2m)} = 0, \quad \hat{a}_{j(2m+1)} = 0,
\end{eqnarray}
for all non-negative values of $m$. However, in the present context, we should note again that the above equations are not valid and are expected to be nonzero. With the above expressions in hand, we now attempt to connect the potential with $a_{jk}$ and start with
decomposing $\Tilde{\xi}$ in real and imaginary parts as follows:
\begin{equation}
    \Tilde{\xi} = \hat{\xi} + i \Breve{\xi}.
\label{eq:tildexi}    
\end{equation}
Therefore, we gather
\begin{equation}
    \hat{\xi} = \sum_{j,k = 0}^{\infty} \hat{a}_{jk} \frac{\rho^j z^k}{(\rho^2 + z^2)^{j+k}}, \quad \Breve{\xi} = \sum_{j,k = 0}^{\infty} \Breve{a}_{jk} \frac{\rho^j z^k}{(\rho^2 + z^2)^{j+k}}.
\label{eq:xi_ajk}    
\end{equation}
By using \ref{eq:ajk}, \ref{eq:tildexi} and \ref{eq:xi_ajk}, we may arrive at the expression
\begin{equation}
\label{eq:F}
    F = \frac{R - \hat{\xi}^2 - \Breve{\xi}^2}{(R^{1/2} + \hat{\xi})^2 + \Breve{\xi}^2},\,\,\,\,\,\,\,\psi = \frac{-2R^{1/2}\Breve{\xi}}{(R^{1/2} + \hat{\xi})^2 + \Breve{\xi}^2},
\end{equation}
where $R = \rho^2 + z^2$. For our purpose, we need to understand the properties of the ${\partial \mathcal{V}_{\rm z}}/{\partial z}$ up to order of $z$, and considering $\mathcal{V}_{\rm z}$ up to order of $z^2$. This would need the knowledge of $g^{tt}, g^{t\phi}, g^{\phi\phi}, F, g_{t\phi}$ up to the order of $z^2$ and $\psi$  up to the order of $z^3$. Therefore, we need contribution from $\hat{a}_{j0}, \hat{a}_{j1}$ and $\Breve{a}_{j0}, \Breve{a}_{j1}$ for our analysis. For the expansion,  we take $z/\rho\ll 1$, and as a result, it is not necessary for $z$ to be very small as long as $z\ll \rho$ is satisfied. Keeping up to the order of $z^3/\rho^3$, we rewrite $\hat{\xi}$ and $\Breve{\xi}$ as follows:
\begin{eqnarray}
\label{eq:xi}
    \hat{\xi} =\sum_{j} \Bigl[\frac{\hat{a}_{j0}}{\rho^j} \{1 - j\frac{z^2}{\rho^2}\} + \frac{\hat{a}_{j1}}{\rho^{j+1}} \{\frac{z}{\rho} - (j+1)\frac{z^3}{\rho^3}\}\Bigr],\nonumber\\
        \Breve{\xi} = \sum_j \Big[\frac{\Breve{a}_{j0}}{\rho^j} \{1 - j\frac{z^2}{\rho^2}\} + \frac{\Breve{a}_{j1}}{\rho^{j+1}} \{\frac{z}{\rho} - (j+1)\frac{z^3}{\rho^3}\}\Big].
\end{eqnarray}
Using the results found in this section, we will find the ${\partial \mathcal{V}_{\rm z}}/{\partial z}$ in the next section.

\section{Results}
\label{results}
In this section, we express the relevant quantities as a series expansion in $z$ and only keep terms up to $z^2$ terms in the potential. Based on our earlier discussions, here, we will derive $\mathcal{V}_1$ for a general stationery, axis-symmetric metric. By substituting \ref{eq:xi} into \ref{eq:F}, we arrive at
\begin{eqnarray}
F &=& \frac{1 + \sum_0^3 g_i z^i/\rho^i}{1 + \sum_0^3 f_i z^i/\rho^i} \equiv \sum_0^3 F_i z^i,\nonumber \\
\psi &=& \frac{\sum_0^3 \psi_i z^i/\rho^i}{\rho^2(1 + \sum_0^3 f_i z^i/\rho^i)} \equiv \sum_{i=0}^3\Psi_i\frac{z^i}{\rho^i},
    \label{eq:F_PSI}
\end{eqnarray}
where, the coefficients can be expressed as follows:
\begin{eqnarray}
f_0 &=&  \sum_j \hat{a}_{j0}\frac{2}{\rho^{j+1}} + \Big[\sum_{j,j'} \frac{1}{\rho^{j+j'+2}} \{\hat{a}_{j0} \hat{a}_{j'0} + \hat{a} \rightarrow \Breve{a}\}\Big], \nonumber \\
f_1 &=& \sum_j \hat{a}_{j1}\frac{2}{\rho^{j+2}} + \Big[\sum_{j,j'} \frac{1}{\rho^{j+j'+3}} 2\{\hat{a}_{j0} \hat{a}_{j'1}+ \hat{a} \rightarrow \Breve{a}\}\Big] \nonumber\\
 f_2&=& 1 + \sum_j -\hat{a}_{j0}\frac{2}{\rho^{j+1}} \left(j -\frac{1}{2}\right)+ \Big[\sum_{j,j'} \frac{1}{\rho^{j+j'+2}}\nonumber \\
& & \hspace{2cm}\{-\hat{a}_{j0} \hat{a}_{j'0} (j + j') +\frac{\hat{a}_{j1}\hat{a}_{j'1}}{\rho^2}+ \hat{a} \rightarrow \Breve{a}\}\Big] \nonumber \\
g_0 &=&  \sum_j \hat{a}_{j0}\frac{2}{\rho^{j+1}} -f_0,\,\,\,\, g_1 = \sum_j \hat{a}_{j1}\frac{2}{\rho^{j+2}} -f_1 \nonumber \\
g_2 &=& \sum_j -\hat{a}_{j0}\frac{2}{\rho^{j+1}} \left(j -\frac{1}{2}\right)+ 2 -f_2 \nonumber \\
\psi_0 &=& \sum_j \Breve{a}_{j0} \frac{2}{\rho^{j-1}},~\psi_2 = \sum_j -\Breve{a}_{j0} \frac{2}{\rho^{j-1}} \left(j-\frac{1}{2}\right),\nonumber \\
\psi_1 &=& \sum_j \Breve{a}_{j1} \frac{2}{\rho^{j}},~\psi_3 = \sum_j -\Breve{a}_{j1} \frac{2}{\rho^{j}} \left(j+\frac{1}{2}\right)
\end{eqnarray}

\begin{eqnarray}
F_0 &=& \frac{1+g_0}{1+f_0},~ F_1 = \frac{g_1(1+f_0)-f_1(1-g_0)}{\big(f_0+1\big){}^2 }, \nonumber \\
F_2 &=& \frac{g_2(1+f_0)^2+(f_2g_0+g_1f_1-f_2)(1+f_0)+f_1^2(1+g_0)}{\big(f_0+1\big){}^3 },\nonumber \\
\Psi_0 &=&  \frac{\psi_0}{\rho^2(1+f_0)},~\Psi_1 =  \frac{\psi_1(1+f_0)-f_1\psi_0}{\rho^2(1+f_0)^2}\nonumber \\
\Psi_2 &=& \frac{(1+f_0)^2\psi_2-(f_1\psi_1+f_2\psi_0)(1+f_0)+f_1^2\psi_0}{\rho^(1+f_0)^3} \\
\Psi_3 &=& \frac{(1+f_0)^3\psi_3 - (1+f_0)^2(f_1\psi_2+f_2\psi_1+f_3\psi_0)}{\rho^2(1+f_0)^4}\\
&+&\frac{(1+f_0)(f_1^2\psi_1+2f_1f_2\psi_0)-\psi_0f_1^3}{\rho^2(1+f_0)^4} \nonumber
\end{eqnarray}
which are functions of $\rho$ only and independent of $z$. 
By employing these expressions, we can obtain the derivatives of the metric components, $g_{tt}$, $g_{t\phi}$, and $g_{\phi \phi}$ [given in \ref{eq:g_tPhi}], which are essential for our study. We start with the following expression for $g_{t\phi}$ and obtain the derivative of $g_{t\phi}$ and $g_{\phi \phi}$
  \begin{equation}
      g_{t\phi}  =  -F  \sum_i  z^i \int I_i d\rho' + N= -F  \sum_i  z^i  \mathcal{I}_i +\sum_i N_i z^i ,
  \end{equation}
\begin{eqnarray}
\label{gtPhi partial}
 \rho^2\frac{\partial g^{t\phi}}{\partial z} &=& -(F_1\mathcal{I}_0-N_1+F_0\mathcal{I}_1) \nonumber\\
 & &-2z(F_1\mathcal{I}_1+F_2\mathcal{I}_0-N_2+F_0\mathcal{I}_2), \nonumber \\ 
 \frac{\partial g^{\phi\phi}}{\partial z} &=& \frac{1}{\rho^2} [F_1 + 2F_2 z],
\end{eqnarray}
where we have used \ref{eq:F_PSI}, and the expressions for $N_1, N_2, I_0, I_1, I_2$ are given as
\begin{eqnarray}
N_1 &=& \frac{\rho \Psi_0'}{F_0}, \nonumber \\
N_2 &=& \frac{\Psi_1'}{2F_0}-\frac{\Psi_1}{2\rho F_0},\\
I_0 &=& \frac{\Psi_1}{F_0^2},\,\,\,I_1 = \frac{2\Psi_2F_0-2F_1\Psi_1\rho}{\rho F_0^3}, \nonumber \\
I_2 &=& \frac{3\Psi_3F_0^2-4F_1\Psi_2 F_0\rho -2 F_2F_0\Psi_1\rho^2 +3F_1^2\Psi_1\rho^2}{F_0^4\rho^2},\nonumber 
\label{eq:N_I}
\end{eqnarray}
where $\Psi_i' = \frac{\partial\Psi_i}{\partial\rho}$. It is now easy to evaluate  the derivative of $g_{tt}$ by using \ref{gtPhi partial} and \ref{eq:N_I}.
~ Finally, we can obtain the expression for ${\partial\mathcal{V}_{\rm z}(z)}/{\partial z}$ by using \ref{eq:Vz} and the derivatives of the metric components. 
 The result is as folllows:
 
 \begin{eqnarray}
 \mathcal{V}_1 &=& -E^2\left(-\frac{\mathcal{I}_0^2 F_1}{\rho^2} + \frac{2\mathcal{I}_0\mathcal{I}_1F_0}{\rho^2} + \frac{F_1}{F_0^2}-\frac{2\mathcal{I}_0N_1 }{\rho^2}+2\frac{\mathcal{I}_0^2F_1}{\rho^2}\right) \nonumber\\ 
 &&-\frac{L_z^2}{\rho^2}F_1+2\frac{EL_z}{\rho^2} (F_0\mathcal{I}_1+F_1\mathcal{I}_0-N_1).\label{eq:V1}
 \end{eqnarray}

 It is easy to notice that in general $\mathcal{V}_1 \neq 0$ from \ref{eq:V1}. The consequence of nonzero $\mathcal{V}_1$ has already been discussed in \ref{Geodesic}. It may be possible that, even though the $z-$ independent term is non-zero for each of the derivatives, in some special, cases they will add up to give a vanishing $z$ independent term. In particular, one may ask whether it is possible to choose conserved energy and momentum in such a way that it would result in $\mathcal{V}_{\rm 1}=0$. Indeed, we find this can be a possibility; however, both the energy and momentum need to be in consonance with $\dot{\rho}=\ddot{\rho}=0$, too, which can put the further restriction in its motion. For example, in the Kerr-NUT spacetime, it is not possible to choose energy and momentum in such a way that it would not only cancel the off-equatorial push but also describe a circular geodesic \cite{Mukherjee:2018dmm}. Therefore, while this is a valid possibility, it does not describe a general outcome of our findings.

For a quick follow-up of the above analysis where the \ES~is respected, we focus on the $z-$independent term. Here, the $z-$independent part depends only on $F_0, F_1, I_0, I_1, N_1$. When \ES~is present, $\Breve{a}_{j(2m)} = 0$ and $\hat{a}_{j(2m+1)} = 0$, hence, $f_1 = g_1 = \psi_0 = \psi_2 = 0$. This implies $F_1 = \Psi_0 = \Psi_2 = I_1 = \mathcal{I}_1 = N_1=0$. Therefore, the $\mathcal{V}_1$ in \ref{eq:V1} vanishes. It is remarkable that in the case of \ES, the $z^0$ terms in every term that constitutes $\frac{\partial \mathcal{V}_z}{\partial z}$ vanish and, as a result, $\mathcal{V}_1$ vanishes. This indicates that the presence of circular equatorial orbit implies the presence of the reflection symmetry.

\section{Observational prospects and possible constraints} \label{observation}
We explore a possible connection between the \ES~and equatorial circular orbits and arrive at the conclusion that in the absence of the equatorial reflection symmetry circular orbit can not exist on the equatorial plane. The parameters which break the reflection symmetry will engineer to elevate the orbit from the equatorial plane and boost it with a force. However, if we assume these parameters are small enough, we can still naively assume the orbits to be equatorial and still perform some astrophysical calculations \cite{Mukherjee:2020how}. Nonetheless, this would introduce a test bed to execute several observational expeditions to confirm the existence of stable orbits and \ES. In fact, the observation of stable circular equatorial orbit will be a telltale signature of the presence of reflection symmetry or a very mild violation of it. For the present purpose, we outline a few such examples where the violation of \ES~may be detectable by observation. For other theoretical and observable impacts, check Refs. \cite{Cunha:2018uzc, Chen:2020aix, Aelst:2020zvf}. 

\subsection{Gravitational wave astronomy}
Consider a binary system prepared in such a way that the spins of the components are either aligned or antialigned with the orbital angular momentum. With time due to the push from $\mathcal{V}_1$, the spins will not stay (anti) aligned even if they were prepared in an (anti) aligned manner. This will result in a nonzero in-plane spin component $(\chi_p)$ (check Ref.~\cite{Schmidt:2014iyl} for the definition). Therefore, for \ES~violating bodies in a binary, $\chi_p$ measurement should be nonzero. Hence, nonzero $\chi_p$ can arise in several different ways. One is due to the formation mechanism of binaries, which has components that respect \ES yet introduce a nonzero $\chi_{p}$ possibly due to spin effects. Another reason for nonzero $\chi_p$ would be due to the absence of \ES. This means that there will be a degeneracy between the formation channel and \ES~violation, and makes it uncertain to arrive at a unique conclusion.

This possible degeneracy can be broken by measuring the multipole moments of a compact object. It is well known that for axis-symmetric bodies there can be two sets of multipole moments: one is the mass moment ($M_{l}$), and another is the current moment ($S_{l}$). For a metric solution with the \ES~like Kerr, both the odd mass and even current moments would identically vanish. However, for a simple illustration of these moments in a Kerr-NUT spacetime which is known to break the \ES, one immediately notices that all the orders for mass and current multipole moments would survive \cite{Mukherjee:2020how}. From an observation perspective, there now exists contemporary tools to measure the quadrupole moment in a binary \cite{Krishnendu:2017shb, Datta:2019euh}. If reflection symmetry is violated, then it is likely that the metric will have nonzero $M_{2l+1}$ and/or $S_{2l}$, i.e., classes of fuzzball solutions \cite{Bena:2020uup, Bena:2020see, Bianchi:2020miz, Bianchi:2020bxa, Bena:2009pyv, Gibbons:2013tqa, Bates:2003vx, Mayerson:2020tpn}. Hence, measuring a nonzero $S_2$ will be a signature of breaking of \ES, along with the nonzero $\chi_p$ observation.
\subsection{Constraints from the accretion disk}
If the metric of these objects does not respect \ES,~then there should be some imprint of such violation on the matter distribution around it. In such cases, depending on the value of $\mathcal{V}_1$, we may expect the matter to be distributed in off-equatorial planes. This, as a result, can give a possible opportunity to constraint $\mathcal{V}_1$ from observation. For example, if we assume that there is a deviation from the reflection symmetry, we gather from \ref{eq:sintheta} that farthest a particle can go from the equatorial plane is $\mathcal{D} \sim |\frac{\mathcal{V}_1}{\omega^2}|$. Depending on the model the radius of an accretion disk around a BH can have radius $r\sim R (\frac{M}{M_{\odot}})$, where $R \sim (10-10^6){\rm Km}$. Traditional thin accretion disks can have scale height $h\sim .01 r$ \cite{Shakura:1972te, frank_king_raine_2002}. Therefore, to ensure a disk structure consistent with most of the observation, we have $\mathcal{D}<h$, which translates to
\begin{eqnarray}
\left \vert \frac{\mathcal{V}_1}{\omega^2} \right \vert< 10^4 {\rm Km} \left(\frac{\zeta}{.01}\right)\left(\frac{R}{10^6{\rm Km}}\right)\left(\frac{M}{M_{\odot}}\right),
\end{eqnarray}
where $\zeta = h/r$. For a given accretion disk, we may be able to constrain the reflection breaking parameters from observation, which may provide a bit of information about the central object. Since the accretion phenomenon is observed with x-ray observations, it requires investigating if it is possible to probe \ES~from the x-ray observations. 
The observed time variability in the x-ray flux emitted
by accreting compact objects (i.e., quasiperiodic oscillations \cite{vanderKlis:2004js, Stella:1998mq, Stella:1999sj, Abramowicz:2001bi}) can possibly shed some light in this regard. Currently, the underlying mechanism is not very well understood, except the belief that they originate from the innermost region of the accretion flow \cite{vanderKlis:2000ca}. Since the physics of accretion disks is very complex, it is challenging to extract accurate information. We will leave such studies for the future.
\section{Conclusion}
\label{conclusion}
We have studied the perturbation of the circular-equatorial orbits of a general stationary axis-symmetric metric in the $z$ direction. In the process, we have identified a set of parameters that are the potential source of the \ES~violation, namely, $\Breve{a}_{j(2m)}$ and $\hat{a}_{j(2m+1)}$. We have shown that, in general, when these parameters are zero (nonzero) \ES~is present (absent), and the small oscillation solution across the equatorial plane is present (absent). This leads us to conclude that the very existence of equatorial circular orbit is an indication that the geometry respects \ES, while it may not be true the way around. To be precise, equatorial circular geodesics may not exist even if the spacetime respects the equatorial reflection symmetry. For example, this can be simply stemmed from the fact that the conserved momentum and energy are not favorable to host any bound equatorial circular geodesic.

In this paper, we argue that in the case of \ES~violation it is unlikely to have a measurement with $\chi_p = 0$. This can be addressed by properly identifying the orbital parameters that will be representative of symmetry violation, and possibly depend on $\Breve{a}_{j(2m)}$ and $\hat{a}_{j(2m+1)}$.
Therefore, to break this stalemate, we need to confront the multipolar structure of the object which would consist of both odd mass multipole moments and even current multipole moments in case the symmetry is violated. By measuring the $S_2$ and $M_3$ components, it would be sufficient to confirm this claim. We also constrained \ES~violating parameters for the stellar and supermassive objects. To our knowledge, this is the first time such a kind of constraint has been found.

\section*{Acknowledgments}
Both the authors are indebted to Sukanta Bose, Sumanta Chakraborty, Naresh Dadhich, Prasun Dhang, Ranjeev Misra, Sanjit Mitra, and Kanak Saha for useful comments and also suggesting changes for the betterment of the article. S.D. would like to thank University Grants Commission (UGC), India, for providing a senior research fellowship, and S.M. is thankful to the Department of Science and Technology, Government of India, for financial support.   

\appendix
\labelformat{section}{Appendix #1} 
\labelformat{subsection}{Appendix #1}
\numberwithin{equation}{section}

\section{Metric components in terms of the Ernst potential}
\label{Ernst_Pot}

In this section, we will derive $g_{t\phi}$ in terms of the Ernst potential. To our knowledge this has not been computed explicitly in the literature. If the metric is stationary and axis symmetric, then there will exist a timelike Killing vector field $\xi^{\alpha}$ and an axial Killing vector field $\mathfrak{A}^{\alpha}$. Then, it is possible to define a vector field $\omega^{\alpha}$, defined as 
\begin{equation}
\label{omega definition}
    \omega_{\alpha} = \epsilon_{\alpha\beta\gamma\delta} \xi^{\alpha}\nabla^{\beta}\xi^{\gamma}
\end{equation}
which satisfies $\nabla_{[\alpha}\omega_{\beta]} = 0$. Therefore, we can define a twist potential $\psi$ as $\omega_{\alpha} = \partial_{\alpha}\psi$.

In the cylindrical coordinate system $(t,\rho,z,\phi)$, the nonzero components can be found as

\begin{eqnarray}
\label{omega result}
\omega_{\rho} &=& -(g_{\rho\rho}g_{zz})^{1/2}g^{zz} \big[g^{t\phi}\partial_{z}g_{tt} + g^{\phi\phi}\partial_z g_{t\phi}\big]\nonumber\\
&=& \frac{1}{\rho}g_{tt}^2 \partial_{z}\left(\frac{g_{t\phi}}{g_{tt}}\right),\nonumber\\
\omega_{z} &=& (g_{\rho\rho}g_{zz})^{1/2}g^{\rho\rho} \big[g^{\phi\phi}\partial_{\rho}g_{t\phi} + g^{t\phi}\partial_{\rho} g_{tt}\big]\nonumber\\
&=& -\frac{1}{\rho}g_{tt}^2 \partial_{\rho}\left(\frac{g_{t\phi}}{g_{tt}}\right).
\end{eqnarray}
From \ref{omega result}, it is simple to find $g_{t\phi}$ as
\begin{equation}
    \frac{g_{t\phi}}{g_{tt}} = \int\frac{\rho}{g_{tt}^2} \partial_{\rho}\psi \,\, dz - \int\frac{\rho}{g_{tt}^2} \partial_{z}\psi \,\, d\rho.
\end{equation}

\section{Example of a \ES~breaking spacetime--Kerr-NUT geometry} \label{Appen:Kerr_NUT}
To display the connection between equatorial symmetry and circular orbits explicitly, we consider an example, namely, Kerr-NUT spacetime, where the reflection symmetry is known to be violated \cite{Mukherjee:2018dmm}. The nonexistence of equatorial circular orbits in \KN~geometry was first claimed in Ref. \cite{Jefremov:2016dpi} and recently explored further in Ref. \cite{Mukherjee:2018dmm}. In the present context, though, we will be more involved in studying the perturbation equation in the theta direction and discuss the near equatorial plane behavior.

The Kerr-NUT spacetime is a solution to vacuum Einstein field equations and describes a stationary, axis-symmetric, and asymptotically nonflat spacetime. To discuss the nature of the angular perturbation, we need to consider the angular geodesic equation in Boyer-Lindquist coordinates $(t,r,\theta,\phi)$,
\begin{eqnarray}
\mathcal{V}_{\theta}=(\dot{\theta})^2=\left(\dfrac{d\theta}{d\tau_{\rm m}}\right)^2 =\Bigl(\lambda \sin^2\theta +(L_{\rm z}-aE)^2 \sin^2\theta  \nonumber \\
-(E P-L_{\rm z})^2-\sin^2\theta (l+a \cos\theta)^2 \Bigr), \nonumber \\
\end{eqnarray}
where $\tau_{\rm m}$ is the Mino time, a dot defines a derivative with respect to it, and $\mathcal{V}_{\theta}$ can be defined as angular potential \cite{Mukherjee:2018dmm}. The quantities $\lambda$, $E$, $L_{\rm z}$, $l$, $a$ are defined as Carter constant, energy, momentum, NUT charge, and angular momentum, respectively, and  $\rho$ and $P$ are given by $\rho^2=r^2+(l+a \cos\theta)^2$ and $P=a\sin^2\theta-2 l \cos\theta$. Let us now assume that the particle is initially confined on the equatorial plane and we expect to study the perturbation originated from the term $\ddot{\theta}$. By setting $\dot{\theta}=0$ and $\theta=\pi/2$, we gather $\lambda=l^2$, which identically vanishes for zero NUT charge, i.e., in the case of Kerr spacetime. In addition to $\dot{\theta}=0$, to have a planner orbit, we need to ensure $\ddot{\theta}=0$, too, which warrants that $\dot{\theta}=0$ remains satisfied along the trajectory. This is where the NUT charge comes into play, and engineers to disobey $\ddot{\theta}=0$. Even then, the $\ddot{\theta}$ equation can be useful to extract information about the variation of $\theta$ near to the equatorial plane, which we will do next. Let us start by introducing the equation
\begin{equation}
\ddot{\theta}=\dfrac{1}{2}\dfrac{d\mathcal{V}(\theta)}{d\theta},
\label{eq:dot_theta}
\end{equation}
which we need to write in terms of the coordinate time, such that it can be useful for an asymptotic observer. To execute this task, we may unfold $\ddot{\theta}$ as
\begin{equation}
\dfrac{d^2\theta}{d \tau^2}=\left(\dfrac{d^2\theta}{dt^2} \right)\left(\dfrac{dt}{d\tau}\right)^2+\left(\dfrac{d\theta}{d\tau}\right)\left(\dfrac{dt}{d\tau}\right)^{-1}\left[\dfrac{d}{dt}\left(\dfrac{dt}{d\tau}\right)\dfrac{dt}{d\tau}\right],
\label{eq:ddot_theta}
\end{equation}
and one easily notices that the second term goes to zero, as we assume $\dot{\theta}=0$ in the first place. Bringing together Eqs. (\ref{eq:dot_theta}) and (\ref{eq:ddot_theta}) and writing $\theta(t)$ as $\theta(t)=\pi/2+\delta \theta(t)$, we arrive at the following expression: 
\begin{equation}
\left(\dfrac{d^2\delta \theta(t)}{dt^2} \right) \left(\dfrac{dt}{d\tau}\right)^2=\dfrac{1}{2}\dfrac{d\mathcal{V}(\theta)}{d\theta}.
\end{equation}
The expression of $\mathcal{U}^t=dt/d\tau$ can be derived from $\mathcal{U}^t=-g^{tt}E+g^{t\phi}L_{\rm z}$, where the metric components are explicitly written in the Appendix. Finally, assuming $\delta \theta(t) \ll 1$ and terms with $\delta \theta(t)^2 $, $\delta \theta(t) \delta \theta ^{\prime}(t)$, $\delta \theta(t)\delta \theta^{\prime \prime}(t)$, and beyond are neglected, we gather
\begin{equation}
\mathcal{C}_3 \delta \theta^{\prime \prime}(t)+\mathcal{C}_2 \delta {\theta}(t)+\mathcal{C}_1=0,
\end{equation}
where a prime denotes a differentiation with respect to $t$. The expressions for $\mathcal{C}_1$, $\mathcal{C}_2$, and $\mathcal{C}_3$ are given as
\begin{eqnarray}
\mathcal{C}_1&=&l \Bigl\{a(2 E^2-1)-2EL_{\rm z}\Bigr\},\nonumber \\
\mathcal{C}_2&=&L_{\rm z}^2+4 E^2 l^2-a^2(E^2-1), \nonumber \\
 \mathcal{C}_3&=&\dfrac{1}{\Delta^2}\bigl\{E(r^2+l^2)^2-2aL_{\rm z}(l^2+Mr)+a^2E\Bigl(r(r+2M)\nonumber \\
 & & \hspace{6cm} +3l^2 \Bigr)\bigr\},
\end{eqnarray}
where $\Delta=r^2-2Mr+a^2-l^2$ becomes zero on the event horizon. The above equation has a generic solution of the form
\begin{equation}
\delta \theta(t)=-\dfrac{\mathcal{C}_1}{\mathcal{C}_2}+A \cos\omega t+B \sin\omega t,
\end{equation}
where $A$ and $B$ are integration constants to be evaluated from the initial conditions and $\omega=\sqrt{\mathcal{C}_2/\mathcal{C}_3}$. It is interesting to point out that $\mathcal{C}_3$ diverges on the horizon, and $\omega$ becomes zero, which no longer describes equatorial timelike circular orbits. This is in consonance with the fact that on the null surface of event horizon no timelike circular orbit can exist. With the initial condition $\delta \theta(t=0)=\delta \theta^{\prime}(t=0)=0$, the above equation turns out to be
\begin{equation}
\delta \theta(t)=-\dfrac{\mathcal{C}_1}{\mathcal{C}_2} \Bigl(1-\cos\omega t\Bigr)=-\dfrac{2\mathcal{C}_1}{\mathcal{C}_2}\sin^2(\omega t/2).
\label{eq:oscillations}
\end{equation}
which may oscillate on either side of $\theta=\pi/2$, depending on the signs of $\mathcal{C}_1$ and $\mathcal{C}_2$. Finally, we should say some words regarding the nature of the oscillations and how it is different from the Kerr case. It is easy to realize that the oscillation solely depends on $\mathcal{C}_1$ and $\mathcal{C}_2$, and among them, $\mathcal{C}_2$ is always positive as far as we are concerned with bound circular geodesics, i.e., $E \leq 1$. Coming to $\mathcal{C}_1$, it can only vanish if we have $l=0$ or find a radius which satisfies $(2 E^2-1)=2EL_{\rm z}$. It turns out that the later option is ruled out as far as one is interested with equatorial circular orbits in NUT spacetime \cite{Mukherjee:2018dmm}, and one is left with no choice but to set $l=0$ to stop the oscillation. Therefore, the NUT charge, which is entirely responsible for breaking the equatorial symmetry, also engineers to angular perturbation. In passing, we should also mention the similar scenario in connection to massless particles. In this case, too, it is possible to arrive at an equation equivalent to \ref{eq:oscillations}, only with the expressions of $\mathcal{C}_{1}$, $\mathcal{C}_{2}$, and $\mathcal{C}_{3}$ changed as follows:
\begin{eqnarray}
\mathcal{C}_1&=&2El \Bigl\{aE-L_{\rm z}\Bigr\},\nonumber \\
\mathcal{C}_2&=&L_{\rm z}^2+ E^2(4l^2-a^2)+l^2, \nonumber \\
 \mathcal{C}_3&=&\dfrac{1}{\Delta^2}\bigl\{E(r^2+l^2)^2-2aL_{\rm z}(l^2+Mr)+a^2E\Bigl(r(r+2M)\nonumber \\
 & & \hspace{6cm} +3l^2 \Bigr)\bigr\},
\end{eqnarray}
\begin{figure}[htp]
\includegraphics[scale=.6]{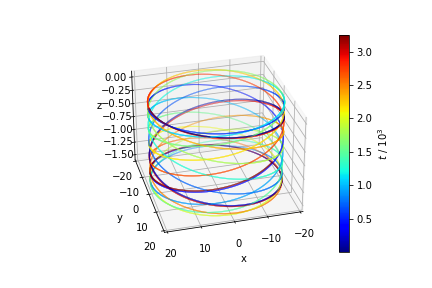}
\caption{In this figure, we demonstrate how a circular orbit in Kerr-NUT spacetime is slowly evolving from the equatorial plane. The radius of the circular orbit is considered to be $r=20M$, and the NUT charge $l=0.1M$. x,y, z, and t are in the units of $M$.}
\label{fig:1}
\end{figure}
\begin{figure}[htp]
\includegraphics[scale=.6]{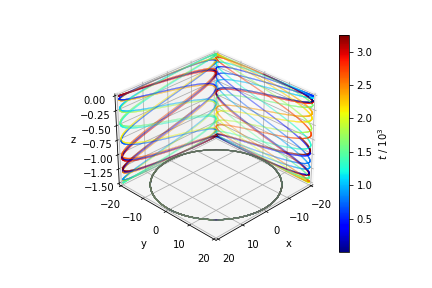}
\caption{In this figure, we demonstrate the evolution by projecting it at constant x, y, and z plane. The radius of the circular orbit is considered to be $r=20M$, and the NUT charge $l=0.1M$. x,y, z, and t are in the units of $M$.}
\label{fig:2}
\end{figure}
The expression for $\mathcal{C}_{1}$ can be set to zero by two possible ways, namely, $l=0$ and $aE=L_{\rm z}$. Like the earlier case, the later option can never give rise to a circular geodesic, and we need to choose $l=0$. Therefore, the massless case is also in agreement with our claim that the NUT charge is solely responsible for having no equatorial circular orbits. For a typical set of parameters, we have shown in \ref{fig:1} and \ref{fig:2} how the NUT charge is influencing a circular orbit which starts from the equatorial plane. Along with time, the NUT charge pushes the orbit from the equatorial plane and the orbit slowly deviates. Note that the general study representing the same for an arbitrary spacetime with no equatorial symmetry is already presented in \ref{results}. For other specific examples check Refs. \cite{Nakashi:2019mvs,Nakashi:2019tbz}.


\bibliographystyle{utphys1.bst}
\bibliography{References.bib}

\providecommand{\href}[2]{#2}\begingroup\raggedright\begin{thebibliography}{10}

\bibitem{Kerr:1963ud}
R.~P. Kerr, ``{Gravitational field of a spinning mass as an example of
  algebraically special metrics},''
  \href{http://dx.doi.org/10.1103/PhysRevLett.11.237}{{\em Phys. Rev. Lett.}
  {\bfseries 11} (1963) 237--238}.

\bibitem{Israel:1967wq}
W.~Israel, ``{Event horizons in static vacuum space-times},''
  \href{http://dx.doi.org/10.1103/PhysRev.164.1776}{{\em Phys. Rev.} {\bfseries
  164} (1967) 1776--1779}.

\bibitem{Wald:1971iw}
R.~M. Wald, ``{Final states of gravitational collapse},''
  \href{http://dx.doi.org/10.1103/PhysRevLett.26.1653}{{\em Phys. Rev. Lett.}
  {\bfseries 26} (1971) 1653--1655}.

\bibitem{Carter:1971zc}
B.~Carter, ``{Axisymmetric Black Hole Has Only Two Degrees of Freedom},''
  \href{http://dx.doi.org/10.1103/PhysRevLett.26.331}{{\em Phys. Rev. Lett.}
  {\bfseries 26} (1971) 331--333}.

\bibitem{Robinson:1975bv}
D.~Robinson, ``{Uniqueness of the Kerr black hole},''
  \href{http://dx.doi.org/10.1103/PhysRevLett.34.905}{{\em Phys. Rev. Lett.}
  {\bfseries 34} (1975) 905--906}.

\bibitem{Cardoso:2017cfl}
V.~Cardoso, E.~Franzin, A.~Maselli, P.~Pani, and G.~Raposo, ``{Testing
  strong-field gravity with tidal Love numbers},''
  \href{http://dx.doi.org/10.1103/PhysRevD.95.089901,
  10.1103/PhysRevD.95.084014}{{\em Phys. Rev.} {\bfseries D95} no.~8, (2017)
  084014}, \href{http://arxiv.org/abs/1701.01116}{{\ttfamily arXiv:1701.01116
  [gr-qc]}}.
[Addendum: Phys. Rev.D95,no.8,089901(2017)].

\bibitem{Sennett:2017etc}
N.~Sennett, T.~Hinderer, J.~Steinhoff, A.~Buonanno, and S.~Ossokine,
  ``{Distinguishing Boson Stars from Black Holes and Neutron Stars from Tidal
  Interactions in Inspiraling Binary Systems},''
  \href{http://dx.doi.org/10.1103/PhysRevD.96.024002}{{\em Phys. Rev.}
  {\bfseries D96} no.~2, (2017) 024002},
\href{http://arxiv.org/abs/1704.08651}{{\ttfamily arXiv:1704.08651 [gr-qc]}}.

\bibitem{Maselli:2017cmm}
A.~Maselli, P.~Pani, V.~Cardoso, T.~Abdelsalhin, L.~Gualtieri, and V.~Ferrari,
  ``{Probing Planckian corrections at the horizon scale with LISA binaries},''
  \href{http://dx.doi.org/10.1103/PhysRevLett.120.081101}{{\em Phys. Rev.
  Lett.} {\bfseries 120} no.~8, (2018) 081101},
\href{http://arxiv.org/abs/1703.10612}{{\ttfamily arXiv:1703.10612 [gr-qc]}}.

\bibitem{Brustein:2020tpg}
R.~Brustein and Y.~Sherf, ``{Quantum Love},''
  \href{http://arxiv.org/abs/2008.02738}{{\ttfamily arXiv:2008.02738 [gr-qc]}}.

\bibitem{Datta:2019euh}
S.~Datta and S.~Bose, ``{Probing the nature of central objects in
  extreme-mass-ratio inspirals with gravitational waves},''
  \href{http://dx.doi.org/10.1103/PhysRevD.99.084001}{{\em Phys. Rev. D}
  {\bfseries 99} no.~8, (2019) 084001},
  \href{http://arxiv.org/abs/1902.01723}{{\ttfamily arXiv:1902.01723 [gr-qc]}}.

\bibitem{Datta:2019epe}
S.~Datta, R.~Brito, S.~Bose, P.~Pani, and S.~A. Hughes, ``{Tidal heating as a
  discriminator for horizons in extreme mass ratio inspirals},''
  \href{http://dx.doi.org/10.1103/PhysRevD.101.044004}{{\em Phys. Rev.}
  {\bfseries D101} no.~4, (2020) 044004},
\href{http://arxiv.org/abs/1910.07841}{{\ttfamily arXiv:1910.07841 [gr-qc]}}.

\bibitem{Datta:2020gem}
S.~Datta, K.~S. Phukon, and S.~Bose, ``{Recognizing black holes in
  gravitational-wave observations: Telling apart impostors in mass-gap
  binaries},'' \href{http://arxiv.org/abs/2004.05974}{{\ttfamily
  arXiv:2004.05974 [gr-qc]}}.

\bibitem{Datta:2020rvo}
S.~Datta, ``{Tidal heating of Quantum Black Holes and their imprints on
  gravitational waves},''
\href{http://arxiv.org/abs/2002.04480}{{\ttfamily arXiv:2002.04480 [gr-qc]}}.

\bibitem{Krishnendu:2017shb}
N.~Krishnendu, K.~Arun, and C.~K. Mishra, ``{Testing the binary black hole
  nature of a compact binary coalescence},''
  \href{http://dx.doi.org/10.1103/PhysRevLett.119.091101}{{\em Phys. Rev.
  Lett.} {\bfseries 119} no.~9, (2017) 091101},
  \href{http://arxiv.org/abs/1701.06318}{{\ttfamily arXiv:1701.06318 [gr-qc]}}.

\bibitem{Cardoso:2016rao}
V.~Cardoso, E.~Franzin, and P.~Pani, ``{Is the gravitational-wave ringdown a
  probe of the event horizon?},''
  \href{http://dx.doi.org/10.1103/PhysRevLett.116.171101}{{\em Phys. Rev.
  Lett.} {\bfseries 116} no.~17, (2016) 171101},
  \href{http://arxiv.org/abs/1602.07309}{{\ttfamily arXiv:1602.07309 [gr-qc]}}.
  [Erratum: Phys.Rev.Lett. 117, 089902 (2016)].

\bibitem{Cardoso:2016oxy}
V.~Cardoso, S.~Hopper, C.~F.~B. Macedo, C.~Palenzuela, and P.~Pani,
  ``{Gravitational-wave signatures of exotic compact objects and of quantum
  corrections at the horizon scale},''
  \href{http://dx.doi.org/10.1103/PhysRevD.94.084031}{{\em Phys. Rev.}
  {\bfseries D94} no.~8, (2016) 084031},
\href{http://arxiv.org/abs/1608.08637}{{\ttfamily arXiv:1608.08637 [gr-qc]}}.

\bibitem{Tsang:2019zra}
K.~W. Tsang, A.~Ghosh, A.~Samajdar, K.~Chatziioannou, S.~Mastrogiovanni,
  M.~Agathos, and C.~Van Den~Broeck, ``{A morphology-independent search for
  gravitational wave echoes in data from the first and second observing runs of
  Advanced LIGO and Advanced Virgo},''
  \href{http://dx.doi.org/10.1103/PhysRevD.101.064012}{{\em Phys.\ Rev.\ D}
  {\bfseries 101} no.~6, (2020) 064012},
  \href{http://arxiv.org/abs/1906.11168}{{\ttfamily arXiv:1906.11168 [gr-qc]}}.

\bibitem{Abedi:2016hgu}
J.~Abedi, H.~Dykaar, and N.~Afshordi, ``{Echoes from the Abyss: Tentative
  evidence for Planck-scale structure at black hole horizons},''
  \href{http://dx.doi.org/10.1103/PhysRevD.96.082004}{{\em Phys. Rev.}
  {\bfseries D96} no.~8, (2017) 082004},
\href{http://arxiv.org/abs/1612.00266}{{\ttfamily arXiv:1612.00266 [gr-qc]}}.

\bibitem{Westerweck:2017hus}
J.~Westerweck, A.~Nielsen, O.~Fischer-Birnholtz, M.~Cabero, C.~Capano, T.~Dent,
  B.~Krishnan, G.~Meadors, and A.~H. Nitz, ``{Low significance of evidence for
  black hole echoes in gravitational wave data},''
  \href{http://dx.doi.org/10.1103/PhysRevD.97.124037}{{\em Phys. Rev.}
  {\bfseries D97} no.~12, (2018) 124037},
\href{http://arxiv.org/abs/1712.09966}{{\ttfamily arXiv:1712.09966 [gr-qc]}}.

\bibitem{Cardoso:2019rvt}
V.~Cardoso and P.~Pani, ``{Testing the nature of dark compact objects: a status
  report},'' \href{http://dx.doi.org/10.1007/s41114-019-0020-4}{{\em Living
  Rev. Rel.} {\bfseries 22} no.~1, (2019) 4},
  \href{http://arxiv.org/abs/1904.05363}{{\ttfamily arXiv:1904.05363 [gr-qc]}}.

\bibitem{Titarchuk:2005rr}
L.~Titarchuk and N.~Shaposhnikov, ``{How to distinguish neutron star and black
  hole x-ray binaries? Spectral index and quasi-periodic oscillation frequency
  correlation},'' \href{http://dx.doi.org/10.1086/429986}{{\em Astrophys. J.}
  {\bfseries 626} (2005) 298--306},
  \href{http://arxiv.org/abs/astro-ph/0503081}{{\ttfamily
  arXiv:astro-ph/0503081}}.

\bibitem{Bambi:2013sha}
C.~Bambi, ``{Measuring the Kerr spin parameter of a non-Kerr compact object
  with the continuum-fitting and the iron line methods},''
  \href{http://dx.doi.org/10.1088/1475-7516/2013/08/055}{{\em JCAP} {\bfseries
  08} (2013) 055}, \href{http://arxiv.org/abs/1305.5409}{{\ttfamily
  arXiv:1305.5409 [gr-qc]}}.

\bibitem{Jiang:2014loa}
J.~Jiang, C.~Bambi, and J.~F. Steiner, ``{Using iron line reverberation and
  spectroscopy to distinguish Kerr and non-Kerr black holes},''
  \href{http://dx.doi.org/10.1088/1475-7516/2015/05/025}{{\em JCAP} {\bfseries
  05} (2015) 025}, \href{http://arxiv.org/abs/1406.5677}{{\ttfamily
  arXiv:1406.5677 [gr-qc]}}.

\bibitem{Bambi:2015kza}
C.~Bambi, ``{Testing black hole candidates with electromagnetic radiation},''
  \href{http://dx.doi.org/10.1103/RevModPhys.89.025001}{{\em Rev. Mod. Phys.}
  {\bfseries 89} no.~2, (2017) 025001},
  \href{http://arxiv.org/abs/1509.03884}{{\ttfamily arXiv:1509.03884 [gr-qc]}}.

\bibitem{Wilkins:1972rs}
D.~C. Wilkins, ``{Bound Geodesics in the Kerr Metric},''
  \href{http://dx.doi.org/10.1103/PhysRevD.5.814}{{\em Phys. Rev. D} {\bfseries
  5} (1972) 814--822}.

\bibitem{o2014geometry}
B.~O'Neill, {\em The geometry of Kerr black holes}.
\newblock Courier Corporation, 2014.

\bibitem{chandrasekhar1998mathematical}
S.~Chandrasekhar, {\em The mathematical theory of black holes}, vol.~69.
\newblock Oxford University Press, 1998.

\bibitem{Jefremov:2016dpi}
P.~Jefremov and V.~Perlick, ``{Circular motion in NUT space-time},''
  \href{http://dx.doi.org/10.1088/0264-9381/33/24/245014}{{\em Class. Quant.
  Grav.} {\bfseries 33} no.~24, (2016) 245014},
  \href{http://arxiv.org/abs/1608.06218}{{\ttfamily arXiv:1608.06218 [gr-qc]}}.
  [Erratum: Class.Quant.Grav. 35, 179501 (2018)].

\bibitem{Mukherjee:2018dmm}
S.~Mukherjee, S.~Chakraborty, and N.~Dadhich, ``{On some novel features of the
  Kerr--Newman-NUT spacetime},''
  \href{http://dx.doi.org/10.1140/epjc/s10052-019-6662-2}{{\em Eur. Phys. J. C}
  {\bfseries 79} no.~2, (2019) 161},
  \href{http://arxiv.org/abs/1807.02216}{{\ttfamily arXiv:1807.02216 [gr-qc]}}.

\bibitem{Chakraborty:2019rna}
C.~Chakraborty and S.~Bhattacharyya, ``{Circular orbits in Kerr-Taub-NUT
  spacetime and their implications for accreting black holes and naked
  singularities},'' \href{http://dx.doi.org/10.1088/1475-7516/2019/05/034}{{\em
  JCAP} {\bfseries 05} (2019) 034},
  \href{http://arxiv.org/abs/1901.04233}{{\ttfamily arXiv:1901.04233
  [astro-ph.HE]}}.

\bibitem{Newman:1963yy}
E.~Newman, L.~Tamubrino, and T.~Unti, ``{Empty space generalization of the
  Schwarzschild metric},''
\href{http://dx.doi.org/10.1063/1.1704018}{{\em J. Math. Phys.} {\bfseries 4}
  (1963) 915}.

\bibitem{LyndenBell:1996xj}
D.~Lynden-Bell and M.~Nouri-Zonoz, ``{Classical monopoles: Newton, NUT space,
  gravimagnetic lensing and atomic spectra},''
  \href{http://dx.doi.org/10.1103/RevModPhys.70.427}{{\em Rev. Mod. Phys.}
  {\bfseries 70} (1998) 427--446},
\href{http://arxiv.org/abs/gr-qc/9612049}{{\ttfamily arXiv:gr-qc/9612049
  [gr-qc]}}.

\bibitem{Audley:2017drz}
{\bfseries LISA} Collaboration, P.~Amaro-Seoane {\em et~al.}, ``{Laser
  Interferometer Space Antenna},''
  \href{http://arxiv.org/abs/1702.00786}{{\ttfamily arXiv:1702.00786
  [astro-ph.IM]}}.

\bibitem{Amaro-Seoane:2014ela}
P.~Amaro-Seoane, J.~R. Gair, A.~Pound, S.~A. Hughes, and C.~F. Sopuerta,
  ``{Research Update on Extreme-Mass-Ratio Inspirals},''
  \href{http://dx.doi.org/10.1088/1742-6596/610/1/012002}{{\em J. Phys. Conf.
  Ser.} {\bfseries 610} no.~1, (2015) 012002},
  \href{http://arxiv.org/abs/1410.0958}{{\ttfamily arXiv:1410.0958
  [astro-ph.CO]}}.

\bibitem{Barack:2003fp}
L.~Barack and C.~Cutler, ``{LISA capture sources: Approximate waveforms,
  signal-to-noise ratios, and parameter estimation accuracy},''
  \href{http://dx.doi.org/10.1103/PhysRevD.69.082005}{{\em Phys. Rev. D}
  {\bfseries 69} (2004) 082005},
  \href{http://arxiv.org/abs/gr-qc/0310125}{{\ttfamily arXiv:gr-qc/0310125}}.

\bibitem{Merritt:2011ve}
D.~Merritt, T.~Alexander, S.~Mikkola, and C.~M. Will, ``{Stellar Dynamics of
  Extreme-Mass-Ratio Inspirals},''
  \href{http://dx.doi.org/10.1103/PhysRevD.84.044024}{{\em Phys. Rev. D}
  {\bfseries 84} (2011) 044024},
  \href{http://arxiv.org/abs/1102.3180}{{\ttfamily arXiv:1102.3180
  [astro-ph.CO]}}.

\bibitem{Apostolatos:1994mx}
T.~A. Apostolatos, C.~Cutler, G.~J. Sussman, and K.~S. Thorne, ``{Spin induced
  orbital precession and its modulation of the gravitational wave forms from
  merging binaries},'' \href{http://dx.doi.org/10.1103/PhysRevD.49.6274}{{\em
  Phys. Rev. D} {\bfseries 49} (1994) 6274--6297}.

\bibitem{Ryan:1995wh}
F.~Ryan, ``{Gravitational waves from the inspiral of a compact object into a
  massive, axisymmetric body with arbitrary multipole moments},''
  \href{http://dx.doi.org/10.1103/PhysRevD.52.5707}{{\em Phys. Rev. D}
  {\bfseries 52} (1995) 5707--5718}.

\bibitem{wald2010general}
R.~M. Wald, {\em General relativity}.
\newblock University of Chicago press, 2010.

\bibitem{fodor1989multipole}
G.~Fodor, C.~Hoenselaers, and Z.~Perj{\'e}s, ``Multipole moments of
  axisymmetric systems in relativity,'' {\em Journal of Mathematical Physics}
  {\bfseries 30} no.~10, (1989) 2252--2257.

\bibitem{Ernst:2006yg}
F.~J. Ernst, V.~S. Manko, and E.~Ruiz, ``{Equatorial symmetry / antisymmetry of
  stationary axisymmetric electrovac spacetimes},''
  \href{http://dx.doi.org/10.1088/0264-9381/23/15/013}{{\em Class. Quant.
  Grav.} {\bfseries 23} (2006) 4945--4952},
  \href{http://arxiv.org/abs/gr-qc/0701112}{{\ttfamily arXiv:gr-qc/0701112}}.

\bibitem{Ernst:2007xq}
F.~J. Ernst, V.~S. Manko, and E.~Ruiz, ``{Equatorial symmetry/antisymmetry of
  stationary axisymmetric electrovac spacetimes. II.},''
  \href{http://dx.doi.org/10.1088/0264-9381/24/9/003}{{\em Class. Quant. Grav.}
  {\bfseries 24} (2007) 2193--2204},
  \href{http://arxiv.org/abs/gr-qc/0701113}{{\ttfamily arXiv:gr-qc/0701113}}.

\bibitem{Mino:2003yg}
Y.~Mino, ``{Perturbative approach to an orbital evolution around a supermassive
  black hole},'' \href{http://dx.doi.org/10.1103/PhysRevD.67.084027}{{\em Phys.
  Rev. D} {\bfseries 67} (2003) 084027},
  \href{http://arxiv.org/abs/gr-qc/0302075}{{\ttfamily arXiv:gr-qc/0302075}}.

\bibitem{Mukherjee:2020how}
S.~Mukherjee and S.~Chakraborty, ``{Multipole moments of compact objects with
  NUT charge: Theoretical and observational implications},''
  \href{http://arxiv.org/abs/2008.06891}{{\ttfamily arXiv:2008.06891 [gr-qc]}}.

\bibitem{Cunha:2018uzc}
P.~V. Cunha, C.~A. Herdeiro, and E.~Radu, ``{Isolated black holes without
  $\mathbb Z_2$ isometry},''
  \href{http://dx.doi.org/10.1103/PhysRevD.98.104060}{{\em Phys. Rev. D}
  {\bfseries 98} no.~10, (2018) 104060},
  \href{http://arxiv.org/abs/1808.06692}{{\ttfamily arXiv:1808.06692 [gr-qc]}}.

\bibitem{Chen:2020aix}
C.-Y. Chen, ``{Rotating black holes without $\mathbb{Z}_2$ symmetry and their
  shadow images},'' \href{http://dx.doi.org/10.1088/1475-7516/2020/05/040}{{\em
  JCAP} {\bfseries 05} (2020) 040},
  \href{http://arxiv.org/abs/2004.01440}{{\ttfamily arXiv:2004.01440 [gr-qc]}}.

\bibitem{Aelst:2020zvf}
K.~Van~Aelst, ``{Note on equatorial geodesics in circular spacetimes},''
  \href{http://dx.doi.org/10.1088/1361-6382/aba80c}{{\em Class. Quant. Grav.}
  {\bfseries 37} no.~20, (2020) 207001},
  \href{http://arxiv.org/abs/2103.01816}{{\ttfamily arXiv:2103.01816 [gr-qc]}}.

\bibitem{Schmidt:2014iyl}
P.~Schmidt, F.~Ohme, and M.~Hannam, ``{Towards models of gravitational
  waveforms from generic binaries II: Modelling precession effects with a
  single effective precession parameter},''
  \href{http://dx.doi.org/10.1103/PhysRevD.91.024043}{{\em Phys. Rev. D}
  {\bfseries 91} no.~2, (2015) 024043},
  \href{http://arxiv.org/abs/1408.1810}{{\ttfamily arXiv:1408.1810 [gr-qc]}}.

\bibitem{Bena:2020uup}
I.~Bena and D.~R. Mayerson, ``{Black Holes Lessons from Multipole Ratios},''
  \href{http://arxiv.org/abs/2007.09152}{{\ttfamily arXiv:2007.09152
  [hep-th]}}.

\bibitem{Bena:2020see}
I.~Bena and D.~R. Mayerson, ``{A New Window into Black Holes},''
  \href{http://arxiv.org/abs/2006.10750}{{\ttfamily arXiv:2006.10750
  [hep-th]}}.

\bibitem{Bianchi:2020miz}
M.~Bianchi, D.~Consoli, A.~Grillo, J.~F. Morales, P.~Pani, and G.~Raposo,
  ``{The multipolar structure of fuzzballs},''
  \href{http://arxiv.org/abs/2008.01445}{{\ttfamily arXiv:2008.01445
  [hep-th]}}.

\bibitem{Bianchi:2020bxa}
M.~Bianchi, D.~Consoli, A.~Grillo, J.~F. Morales, P.~Pani, and G.~Raposo,
  ``{Distinguishing fuzzballs from black holes through their multipolar
  structure},'' \href{http://arxiv.org/abs/2007.01743}{{\ttfamily
  arXiv:2007.01743 [hep-th]}}.

\bibitem{Bena:2009pyv}
I.~L.~R. Bena, {\em {Black Holes, Black Rings and their Microstates}}.
\newblock PhD thesis, Saclay, 2009.

\bibitem{Gibbons:2013tqa}
G.~Gibbons and N.~Warner, ``{Global structure of five-dimensional fuzzballs},''
  \href{http://dx.doi.org/10.1088/0264-9381/31/2/025016}{{\em Class. Quant.
  Grav.} {\bfseries 31} (2014) 025016},
  \href{http://arxiv.org/abs/1305.0957}{{\ttfamily arXiv:1305.0957 [hep-th]}}.

\bibitem{Bates:2003vx}
B.~Bates and F.~Denef, ``{Exact solutions for supersymmetric stationary black
  hole composites},'' \href{http://dx.doi.org/10.1007/JHEP11(2011)127}{{\em
  JHEP} {\bfseries 11} (2011) 127},
  \href{http://arxiv.org/abs/hep-th/0304094}{{\ttfamily arXiv:hep-th/0304094}}.

\bibitem{Mayerson:2020tpn}
D.~R. Mayerson, ``{Fuzzballs and Observations},''
  \href{http://arxiv.org/abs/2010.09736}{{\ttfamily arXiv:2010.09736
  [hep-th]}}.

\bibitem{Shakura:1972te}
N.~I. Shakura and R.~A. Sunyaev, ``{Black holes in binary systems.
  Observational appearance},'' {\em Astron. Astrophys.} {\bfseries 24} (1973)
  337--355.

\bibitem{frank_king_raine_2002}
J.~Frank, A.~King, and D.~Raine,
  \href{http://dx.doi.org/10.1017/CBO9781139164245}{{\em Accretion Power in
  Astrophysics}}.
\newblock Cambridge University Press, 3~ed., 2002.

\bibitem{vanderKlis:2004js}
M.~van~der Klis, ``{A Review of rapid x-ray variability in x-ray binaries},''
  \href{http://arxiv.org/abs/astro-ph/0410551}{{\ttfamily
  arXiv:astro-ph/0410551}}.

\bibitem{Stella:1998mq}
L.~Stella and M.~Vietri, ``{Khz quasi periodic oscillations in low mass x-ray
  binaries as probes of general relativity in the strong field regime},''
  \href{http://dx.doi.org/10.1103/PhysRevLett.82.17}{{\em Phys. Rev. Lett.}
  {\bfseries 82} (1999) 17--20},
  \href{http://arxiv.org/abs/astro-ph/9812124}{{\ttfamily
  arXiv:astro-ph/9812124}}.

\bibitem{Stella:1999sj}
L.~Stella, M.~Vietri, and S.~Morsink, ``{Correlations in the qpo frequencies of
  low mass x-ray binaries and the relativistic precession model},''
  \href{http://dx.doi.org/10.1086/312291}{{\em Astrophys. J. Lett.} {\bfseries
  524} (1999) L63--L66},
  \href{http://arxiv.org/abs/astro-ph/9907346}{{\ttfamily
  arXiv:astro-ph/9907346}}.

\bibitem{Abramowicz:2001bi}
M.~A. Abramowicz and W.~Kluzniak, ``{A Precise determination of angular
  momentum in the black hole candidate GRO J1655-40},''
  \href{http://dx.doi.org/10.1051/0004-6361:20010791}{{\em Astron. Astrophys.}
  {\bfseries 374} (2001) L19},
  \href{http://arxiv.org/abs/astro-ph/0105077}{{\ttfamily
  arXiv:astro-ph/0105077}}.

\bibitem{vanderKlis:2000ca}
M.~van~der Klis, ``{Millisecond oscillations in x-ray binaries},''
  \href{http://dx.doi.org/10.1146/annurev.astro.38.1.717}{{\em Ann. Rev.
  Astron. Astrophys.} {\bfseries 38} (2000) 717--760},
  \href{http://arxiv.org/abs/astro-ph/0001167}{{\ttfamily
  arXiv:astro-ph/0001167}}.

\bibitem{Nakashi:2019mvs}
K.~Nakashi and T.~Igata, ``{Innermost stable circular orbits in the
  Majumdar-Papapetrou dihole spacetime},''
  \href{http://dx.doi.org/10.1103/PhysRevD.99.124033}{{\em Phys. Rev. D}
  {\bfseries 99} no.~12, (2019) 124033},
  \href{http://arxiv.org/abs/1903.10121}{{\ttfamily arXiv:1903.10121 [gr-qc]}}.

\bibitem{Nakashi:2019tbz}
K.~Nakashi and T.~Igata, ``{Effect of a second compact object on stable
  circular orbits},'' \href{http://dx.doi.org/10.1103/PhysRevD.100.104006}{{\em
  Phys. Rev. D} {\bfseries 100} no.~10, (2019) 104006},
  \href{http://arxiv.org/abs/1908.10075}{{\ttfamily arXiv:1908.10075 [gr-qc]}}.

\end{thebibliography}\endgroup

\end{document}